\documentclass[a4paper, 12pt]{article}

\usepackage[T1]{fontenc}
\usepackage{graphicx}
\usepackage[margin=2cm]{geometry}
\usepackage[affil-it]{authblk}
\usepackage[caption=false,singlelinecheck=false]{subfig}
\usepackage{float}
\usepackage{booktabs}
\usepackage{ulem}
\usepackage{amsmath}
\usepackage{amssymb}
\usepackage{amsfonts}
\usepackage{amsthm}
\usepackage{nicefrac}
\usepackage{newtxtext}
\usepackage{newtxmath}
\usepackage{slashed}
\usepackage{enumitem}
\usepackage{multirow}
\usepackage{rotating}
\usepackage{cite}
\usepackage{mathrsfs}
\usepackage{anyfontsize}
\usepackage{placeins} \usepackage[colorlinks, urlcolor=black,
  citecolor=blue, link color=black]{hyperref}

\newcommand*{\DOI}[2]{\href{https://doi.org/\detokenize{#1}}{#2}}

\newcommand{\modulus}[1]{\ensuremath{\left| #1 \right|}}

\renewcommand{\Re}{\textrm{Re}}
\newcommand{\dd}{\ensuremath{\mathrm{d}}}

\numberwithin{equation}{section}

\allowdisplaybreaks

\begin{document}
\title{\bf Phenomenological explorations of a \\ quasi-degenerate \texorpdfstring{$\boldsymbol{Z'}$}{Z'} boson}

\author[1,2]{C.\ S.\ Kim\thanks{cskim@yonsei.ac.kr}}
\author[1]{Donghun Lee\thanks{ehdgns6337@naver.com}}
\author[1]{Gyeonghyun Min\thanks{mk9538@naver.com}}
\author[3]{Sechul Oh\thanks{scohph@yonsei.ac.kr}}
\author[1]{Dibyakrupa Sahoo\thanks{Remote Research Fellow, sdibyakrupa@gmail.com}}

\affil[1]{\normalsize Department of Physics and IPAP, Yonsei University, Seoul 03722, Korea}%

\affil[2]{\normalsize CPNR, Department of Physics, Chonnam National University, Gwangju 61186, Korea}%

\affil[3]{\normalsize University College, Yonsei University, Incheon 21983, Korea}%

\maketitle

\begin{abstract}
We do a general model-independent phenomenological analysis of
the curious case of a quasi-degenerate $Z'$ boson, i.e.\ a $Z'$ boson
whose mass is close to that of the Standard Model (SM) $Z$ boson, with
SM $Z$-like fermionic couplings. Assuming that it exists hidden or
unresolved in the present experimental data, we show that the
experimentally measured invisible width can constrain the coupling of
such a $Z'$ to SM fermions. From tree-level calculation alone we show that
presence of such a $Z'$ boson affects both the cross-section and
forward-backward asymmetry of the process $e^+ e^- \to f \bar{f}$
where $f$ (with $f\neq e^-$) can be a lepton or quark. We also show
that the dimensionless difference-over-sum (DoS) ratios of both
cross-section and forward-backward asymmetry are more effective
observables to probe the $Z'$-pole.
\end{abstract}

\section{Introduction}\label{sec:intro}

Across a large number of theoretically and phenomenologically
motivated proposals on extensions of the Standard Model (SM) of
particle physics, an additional exotic electrically neutral,
colorless, massive spin-1 gauge boson usually denoted as $Z'$ is
hypothesized to exist, see e.g.\ the reviews
\cite{Leike:1998wr,Langacker:2008yv,ParticleDataGroup:2026mpi} and
references therein. Despite many past and ongoing experimental
searches for it \cite{CDF:2005hun, CDF:2006uji, CDF:2008ieg,
D0:2010kuq, CDF:2010zwq, CDF:2011nxq, CDF:2011pih, ATLAS:2019erb,
ATLAS:2019fgd, CMS:2019buh, ATLAS:2020zms, LHCb:2020ysn, CMS:2021ctt,
ATLAS:2021suo, ATLAS:2021ldb, CMS:2022fsw, ATLAS:2023vxg,
ATLAS:2023taw, CMS:2023ooo, ATLAS:2023fds, CMS:2023byi, CMS:2023hwl,
ATLAS:2023kao, ATLAS:2024qvg, ATLAS:2024uvu, ATLAS:2024qqm,
CMS:2024zqs, ATLAS:2024bms, CMS:2025wqy, CMS:2025xvx, CMS:2026fjw,
CMS:2026lsc, CMS:2026gdz}, we do not have any clear, unambiguous,
direct signals of its existence at present. A general trend in the
experimental studies so far has been to push the lower limit on
possible mass of the $Z'$ to ever increasing values. Nevertheless, it
is an interesting question to ponder whether the $Z'$ might be lying
within the experimental energy ranges searched so far, especially in
the neighbourhood of the SM $Z$ boson. As some recent findings suggest
\cite{Qiu:2023zfr, LoChiatto:2024guj}, such a quasi-degenerate  $Z'$
boson (i.e.\ a $Z'$ which is practically almost degenerate with the SM
$Z$ boson) is very much a viable possibility worth exploring in future
experiments. Such theoretical possibilities were also explored earlier
in Refs.~\cite{Hook:2010tw, Dermisek:2011xu, Dermisek:2012qx}.
However, the only experimental paper \cite{L3:1993zvy}, in our
knowledge, probing such a possibility was essentially motivated by
possible quarkonium states which could affect the $Z$-pole
characteristics \cite{Franzini:1985nt, Hall:1985jf}.

In this paper we concern ourselves with finding effective answers to
the following set of phenomenologically relevant questions on
detecting and characterizing such a unique quasi-degenerate $Z'$
possibility:
\begin{enumerate}
\item Since the production and detection of the $Z'$ would involve the
strength of its couplings to SM fermions (say, `$f$'), what would be
the most general structure of a renormalizable $Z'ff$ interaction
Lagrangian? We aim to keep our analysis model-independent, i.e.\ not
dwell on any specific gauge group extension of the SM.

\item What are the useful experimental observables that can probe such
a quasi-degenerate $Z'$ possibility? We specifically confine our
analysis to production of the $Z'$ from $e^+ e^-$ annihilation, and
ask how those observables get affected by taking into account the
interference from photon ($\gamma$), $Z$ and $Z'$ mediated diagrams?

\item Keeping our analysis simple, by considering only tree-level
contributions, we ask whether the parameters related to such a $Z'$
can be constrained from existing experimental results? We also discuss
some interesting possibilities which can evade existing experimental
constraints and how can they possibly be constrained from future
studies.
\end{enumerate}

Our article is structured as follows. In Sec.~\ref{sec:Lagrangian} we
discuss the model-independent most general $Z'ff$ interaction
Lagrangian. Immediately we follow this up by analyzing various
phenomenologically relevant observables in Sec.~\ref{sec:pheno}. This
is followed by a numerical study and discussions in
Sec.~\ref{sec:numerical-discussion}. Finally we conclude in
Sec.~\ref{sec:conclusion} highlighting the important aspects of our
analysis.

\section{Interaction of \texorpdfstring{$\boldsymbol{Z'}$}{Z'} with SM fermions}\label{sec:Lagrangian}

For a direct production of the $Z'$ boson as well as its detection in
any collider experiment, the $Z'$ must interact with the SM fermions.
Imposing Lorentz invariance alone and not restricting oneself to
any specific UV-complete gauge theory at all, suggests the following
structure for the interaction Lagrangian describing all possible
interactions of the SM fermions with $Z'$,
\begin{align}
\mathscr{L}_{Z'f\bar{f}} &= -i \, \sum_f \left[\left(\partial_\mu
\overline{\psi_{f}}\right) \, \left(\mathsf{g}_{S1}^{\prime f} \mathbf{1}
+ \mathsf{g}_{P1}^{\prime f} \gamma^5 \right) \, \psi_{f} +
\overline{\psi_{f}} \, \left(\mathsf{g}_{S2}^{\prime f} \mathbf{1} +
\mathsf{g}_{P2}^{\prime f} \gamma^5 \right) \, \left(\partial_\mu
\psi_{f}\right)\right] \, Z^{\prime\mu} \nonumber\\%
&\quad - i \, \sum_f \left[ \overline{\psi_{f}} \, \gamma_{\mu}
\left(\mathsf{g}_{V}^{\prime f} \mathbf{1} - \mathsf{g}_A^{\prime f}
\gamma^5 \right) \, \psi_{f} \right] \, Z^{\prime\mu} - i \, \sum_f
\left[ \overline{\psi_{f}} \, \sigma_{\mu\nu} \, \left(
\mathsf{g}_{TT}^{\prime f} \mathbf{1} + \mathsf{g}_{PT}^{\prime f}
\gamma^5 \right) \, \psi_{f} \right] \, F_{Z^{\prime}}^{\mu\nu}
\nonumber\\%
&\quad -i \, \sum_f \left[\left(\partial^\mu
\overline{\psi_{f}}\right) \, \sigma_{\mu\nu}
\left(\mathsf{g}_{TT1}^{\prime f} \mathbf{1} + \mathsf{g}_{PT1}^{\prime
f} \gamma^5 \right) \, \psi_{f} + \overline{\psi_{f}} \,
\sigma_{\mu\nu} \left(\mathsf{g}_{TT2}^{\prime f} \mathbf{1} +
\mathsf{g}_{PT2}^{\prime f} \gamma^5 \right) \, \left(\partial^\mu
\psi_{f}\right)\right] \, Z^{\prime\nu} . \label{eq:gen-Lag-full}
\end{align}
where $\psi_f$ denotes the fermion field corresponding to fermion $f$
(which can be any quarks and leptons of the SM, as encoded in the
summation over $f$), $\mathsf{g}_{Si}^{\prime f}$,
$\mathsf{g}_{Pi}^{\prime f}$, $\mathsf{g}_{V}^{\prime f}$,
$\mathsf{g}_{A}^{\prime f}$, $\mathsf{g}_{TT(i)}^{\prime f}$ and
$\mathsf{g}_{PT(i)}^{\prime f}$ for $i=1,2$ are called the scalar,
pseudo-scalar, vector, axial-vector, `true' tensor and `pseudo'-tensor
coupling constants respectively, $\mathbf{1}$ is the $4\times 4$
identity matrix, and $F_{Z'}^{\mu\nu}$ is the field-strength tensor
for $Z'$ defined as $F_{Z'}^{\mu\nu} = \partial^{\mu} Z^{\prime \nu} -
\partial^{\nu} Z^{\prime \mu}$. Note that we are considering no
tree-level violations of flavor in Eq.~\eqref{eq:gen-Lag-full}, e.g.\
there are no vertices of the type $Z' \ell \bar{\ell}'$ or $Z' q
\bar{q}'$ for leptons $\ell \neq \ell'$ and quarks $q \neq q'$. We
also note that $\mathsf{g}_{Si}^{\prime f}$, $\mathsf{g}_{Pi}^{\prime
f}$, $\mathsf{g}_{TT(i)}^{\prime f}$ and $\mathsf{g}_{PT(i)}^{\prime
f}$ for $i=1,2$ are dimensionful coupling constants with mass
dimension $-1$, while renormalizability criteria requires the coupling
constants to be dimensionless. Therefore, if we impose Lorentz
invariance and renormalizability, only vector and axial-vector
interactions of $Z'$ with SM fermions is allowed\footnote{In this
article we consider the light active neutrinos as Dirac fermions. But
one might consider the neutrinos to be Majorana fermions. Since a
Majorana neutrino is its own antineutrino, by definition, one would
need to ensure quantum statistical antisymmetrization for the
appropriate terms in the Lagrangian. If we were to consider Majorana
neutrinos, then the Lorentz invariant, renormalizable interaction
Lagrangian $\mathscr{L}_{Z'\nu\bar{\nu}}$ would have the axial-vector
interaction alone, as discussed in \cite{Kim:2022xjg,
Denner:1992vza}.}. Thus, the most general model-independent, Lorentz
invariant and renormalizable Lagrangian for the interaction of $Z'$
with SM fermions, assuming no lepton and quark flavor violations at
the $Z' f \bar{f}$ vertex, is given by
\begin{equation}\label{eq:gen-Lag-Ren}
\mathscr{L}_{Z'f\bar{f}} = - i \, \frac{g_{Z^{\prime}}}{2} \, \sum_f
\left[ \overline{\psi_{f}} \, \gamma_{\mu} \left(g_{V}^{\prime f}
\mathbf{1} - g_A^{\prime f} \gamma^5 \right) \, \psi_{f} \right] \,
Z^{\prime\mu},
\end{equation}
where we have introduced two new vector and axial-vector coupling
constants $g_V^{\prime f}$ and $g_A^{\prime f}$ and here $g_{Z'}$ is
the new neutral-current coupling constant, such that
\begin{equation}\label{eq:gV-gA}
\mathsf{g}_V^{\prime f} = \left(\tfrac{1}{2} g_{Z'} \right)
g_V^{\prime f}, \qquad \mathsf{g}_A^{\prime f} = \left(\tfrac{1}{2}
g_{Z'} \right) g_A^{\prime f}.
\end{equation}
Although Eq.~\eqref{eq:gen-Lag-Ren} seems to suggest that we have
three independent parameters:  $g_V^{\prime f}$, $g_A^{\prime f}$ and
$g_{Z'}$, only the combinations $\mathsf{g}_V^{\prime f}$ and
$\mathsf{g}_A^{\prime f}$, as given in Eq.~\eqref{eq:gV-gA}, are the
two \textit{independent phenomenological parameters} in $Z'f \bar{f}$
couplings\footnote{ In specific UV-complete $Z'$ models, it would be
possible to have all the three coupling constants $g_V^{\prime f}$,
$g_A^{\prime f}$ and $g_{Z'}$ as independent parameters. However, in a
model-independent phenomenological analysis without introducing
specific gauge group like us, only the combinations
$\mathsf{g}_V^{\prime f}$ and $\mathsf{g}_A^{\prime f}$ are
independent parameters.}, and their physical relevance is clear as
shown in Eq.~\eqref{eq:kappa_V_A} and onwards. Phenomenologically, one
can consider $g_{Z'}/2$ as an overall constant factor common to all
the vector and axial-vector coupling constants for all the fermions,
as explicitly evident from Eq.~\eqref{eq:gV-gA}.

The resemblance of Eq.~\eqref{eq:gen-Lag-Ren} with the following part
of the SM Lagrangian describing interaction of the SM $Z$ boson with
SM fermions is completely intentional,
\begin{equation}\label{eq:L-Z2ff}
\mathscr{L}^{\textrm{SM}}_{Zf\bar{f}} = - i\, \frac{g_Z}{2} \sum_f
\left[ \overline{\psi}_f \, \gamma^\mu \left(g_V^f \mathbf{1} - g_A^f
\gamma^5\right) \psi_f \right] \, Z_\mu,
\end{equation}
where the neutral current weak coupling constant $g_Z$ is given by
$g_Z = g_W/\cos\theta_W$, with $g_W$ being the weak coupling constant
($g_W = e/\sin\theta_W$, where $e$ is the electric charge of
positron), $\theta_W$ being the Weinberg angle (also called the weak
mixing angle). In this article we assume that the coupling constants
$g_V^{\prime f}$ and $g_A^{\prime f}$ are real. Thus our
model-independent analysis respects CP symmetry.

It is important to note that the presence of both vector and
axial-vector couplings do not necessarily require extending the SM
gauge group by an additional $SU(2)$ group. For example, they can
arise even in presence of an extra $U(1)'$ group under which the
left-chiral and right-chiral fermions have different $U(1)'$ charges.
Moreover, consideration of kinetic mixing of $Z$ and $Z'$ can also
contribute to vector and axial-vector couplings of
Eq.~\eqref{eq:gen-Lag-Ren} via appropriate redefinition of the fields.
As far as our phenomenological study is concerned, it suffices to note
that Lorentz invariance and renormalizability are necessary and
sufficient for writing Eq.~\eqref{eq:gen-Lag-Ren}. It also helps our
analysis remain agnostic to and independent of any specific new
physics (NP) model.

An `equivalent' set of independent parameters can be introduced by
comparison with SM $Zf\bar{f}$ couplings of Eq.~\eqref{eq:L-Z2ff} as
follows,
\begin{equation}\label{eq:kappa_V_A}
\kappa_V^f \equiv \frac{g_{Z'} \, g_V^{\prime f}}{g_Z \, g_V^f},
\qquad \kappa_A^f \equiv \frac{g_{Z'} \, g_A^{\prime f}}{g_Z \,
g_A^f}.
\end{equation}
Using Eq.~\eqref{eq:kappa_V_A} we can rewrite
Eq.~\eqref{eq:gen-Lag-Ren} as follows,
\begin{equation}\label{eq:gen-Lag-Zpff-v1}
\mathscr{L}_{Z'f\bar{f}} = - i \left(\frac{g_Z}{2} \right) \, \sum_f
\left[ \overline{\psi_{f}} \, \gamma_{\mu} \left(\kappa_V^f \,
g_{V}^{f} \, \mathbf{1} - \kappa_A^f \, g_A^{f} \, \gamma^5 \right) \,
\psi_{f} \right] \, Z^{\prime \mu}.
\end{equation}
If $\kappa_A^f = 0$ (or $\kappa_V^f = 0$), then the $Z'$ has `purely
vector' (or `purely axial-vector') coupling. To simplify our
phenomenological analysis, without sacrificing the necessary rigor to
study the observables later, we make the following two simplifying
assumptions:
\begin{enumerate}
\item \textit{The scaling parameters $\kappa_V^f$ and $\kappa_A^f$ are
universal for all SM fermions}, i.e.\
\begin{equation}\label{eq:kappa_V_kappa_A}
\kappa_V \equiv \kappa_V^f, \qquad \kappa_A \equiv \kappa_A^f.
\end{equation}
\item \textit{The vector and axial-vector couplings are scaled
similarly}, i.e.\
\begin{equation}\label{eq:kappa}
\kappa \equiv \kappa_V = \kappa_A.
\end{equation}
\end{enumerate}
With these simplifying assumptions, Eq.~\eqref{eq:gen-Lag-Zpff-v1} is
simplified to
\begin{equation}\label{eq:gen-Lag-Zpff-v2}
\mathscr{L}_{Z'f\bar{f}} = - i \, \kappa \, \left(\frac{g_Z}{2} \, 
\right) \, \sum_f \left[ \overline{\psi_{f}} \, \gamma_{\mu}
\left(g_{V}^{f} \, \mathbf{1} - g_A^{f} \, \gamma^5 \right) \,
\psi_{f} \right] \, Z^{\prime \mu}.
\end{equation}
Since Eq.~\eqref{eq:gen-Lag-Zpff-v2} essentially differs from
Eq.~\eqref{eq:L-Z2ff} by the overall scaling factor $\kappa$ and
replacing the $Z^{\mu}$ field by $Z^{\prime \mu}$, we have only two NP
parameters: the \textit{scaling factor} $\kappa$ and the \textit{mass
of the $Z'$ boson} $m_{Z'}$. Note that the choice $\kappa=1$ exactly
corresponds to the popular phenomenological case referred to as
\textit{sequential Standard Model} (SSM) in the literature. And if we
put $\kappa=0$, then the $Z'$ would have no interaction with SM
fermions implying no collider access to such NP possibility.
Therefore, all interesting NP possibilities in our model-independent
analysis would correspond to $0 < \kappa \leqslant 1$. Additionally,
we shall consider $m_{Z'}$ in the immediate neighbourhood of $m_Z$,
the mass of the SM $Z$ boson,
\begin{equation}\label{eq:mZp-range}
m_Z - n \, \Gamma_Z \leqslant m_{Z'} \leqslant m_Z + n \, \Gamma_Z,
\end{equation}
with $\Gamma_Z$ being the total decay width of the SM $Z$
boson\footnote{In this article we use the following values as reported
by the particle data group (PDG)~\cite{ParticleDataGroup:2026mpi},
$m_{Z} = 91.1879 \pm 0.0020$~GeV and $\Gamma_{Z} = 2.4955 \pm
0.0023$~GeV.}, and $n\leqslant 1$ corresponds to the quasi-degenerate
$Z'$ case. Later in the article, for the sole purpose of illustration
alone, we choose $m_{Z'}= m_Z \pm 0.8 \Gamma_Z$ as representatives of
the various possibilities for $m_{Z'}$. Thus,
Eq.~\eqref{eq:gen-Lag-Zpff-v2} together with the conditions $0 <
\kappa \leqslant 1$ and the one given in Eq.~\eqref{eq:mZp-range}
constitute our phenomenological basis.

\section{Phenomenological Observables}\label{sec:pheno}

For our phenomenological study we assume that the $Z'$ boson couples
to SM fermions alone, such that its total decay width is equal to the
sum of its partial decay widths to all the SM fermion pairs, i.e.\
$\Gamma_{Z'} = \sum_f \Gamma_{Z' \to f\bar{f}}$, where $f$ is a SM
fermion. This assumption predicts a conservative minimum value for
$\Gamma_{Z'}$, if the $Z'$ were to couple with any hithertho unknown
non-SM exotic fermions and if its decay to such fermion pairs is
kinematically allowed. Since we are purely working with a picture
where the SM particle content is enhanced by an additional $Z'$ alone
and we make no assumptions about specific gauge structure extension of
the SM, we do not have any exotic fermions in our picture. Below, we
investigate how the invisible decay widths of $Z$ and $Z'$, differential
scattering cross-section and specifically forward-backward asymmetry
can be used to probe the presence of $Z'$.

\subsection{Invisible decay widths of \texorpdfstring{$\boldsymbol{Z}$}{Z} and \texorpdfstring{$\boldsymbol{Z'}$}{Z'}}\label{subsec:pheno-decay_widths}

Starting from Eq.~\eqref{eq:gen-Lag-Zpff-v2} and neglecting all
effects from SM fermion masses\footnote{Here we are assuming that
$m_{Z'} \approx m_Z$ such that $Z' \to t\bar{t}$ decay is
kinematically forbidden. All other fermion pairs are
kinematically allowed. In the quasi-degenerate $Z'$ case, all other
fermion masses can be safely neglected with respect to $m_{Z}$ and
$m_{Z'}$.} we find that
\begin{equation}\label{eq:Gamma_Zp}
\Gamma_{Z'} = \kappa^2 \, \frac{G_F \, m_Z^2 \, m_{Z'}}{6\sqrt{2}\pi}
\, \sum_f N_c^f \, \left( \big(g_V^{f}\big)^2 + \big(g_A^{f}\big)^2
\right),
\end{equation}
where $G_F/\sqrt{2} = g_Z^2/(8 m_Z^2)$ is used to replace $g_Z$ in
terms of the Fermi coupling constant $G_F$, and $N_c^f$ is the color
factor for fermion $f$ (it is equal to $3$ for quarks and it is $1$
for leptons). It is useful to compare Eq.~\eqref{eq:Gamma_Zp} with the
corresponding expression for decay rate of the SM $Z$ boson,
\begin{equation}\label{eq:Gamma_Z}
\Gamma_{Z} = \frac{G_F \, m_Z^3}{6\sqrt{2} \pi} \sum_f N_c^f \, \left(
\big(g_V^{f}\big)^2 + \big(g_A^{f}\big)^2 \right),
\end{equation}
which implies that
\begin{equation}\label{eq:Gamma_Z_Zp_relation}
\Gamma_{Z'} = \kappa^2 \left( m_{Z'} / m_Z \right) \Gamma_{Z}.
\end{equation}

\begin{figure}[hb]
\centering%
\includegraphics[scale=0.825]{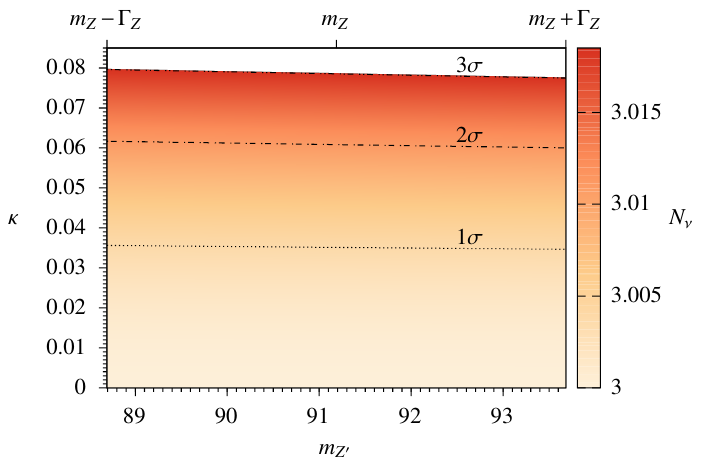}%
\caption{The physically allowed region for $\kappa$ corresponds to the
experimentally measured value $N_{\nu}$, as shown in Eq. (3.5).}%
\label{fig:kappa-mZp-Nnu}
\end{figure}

In the limit of considering the $Z'$ to be practically degenerate with
the SM $Z$ boson, we note that the experimentally inferred invisible
width of the $Z$ boson ($\Gamma_{\textrm{inv}}^{\textrm{exp}}$) could
be contaminated from the presence of the near by $Z'$ decaying to SM
neutrinos,
\begin{equation}\label{eq:Gamma-inv-exp}
\Gamma_{\textrm{inv}}^{\textrm{exp}} = \sum_{\ell=e,\mu,\tau} \left(
\Gamma_{Z \to \nu_{\ell} \bar{\nu}_{\ell}}^{\textrm{SM}} + \Gamma_{Z'
\to \nu_{\ell} \bar{\nu}_{\ell}}^{\textrm{NP}} \right) = 3 \,
\Gamma_{Z \to \nu_{\ell} \bar{\nu}_{\ell}}^{\textrm{SM}} \Big(1 +
\kappa^2 \, m_{Z'} / m_Z \Big),
\end{equation}
which implies that the experimental observable $N_\nu$ used as an
estimate of the number of light neutrinos is given by,
\begin{equation}\label{eq:N_nu-ZZp}
N_{\nu} \equiv \Gamma_{\textrm{inv}}^{\textrm{exp}} / \Gamma_{Z \to
\nu_\ell \bar\nu_\ell}^{\textrm{SM}} = 3 \, \left(1 + \kappa^2 \,
m_{Z'} / m_Z \right).
\end{equation}
Therefore, the experimentally measured value $N_{\nu} = 2.9963 \pm
0.0074$~\cite{ParticleDataGroup:2026mpi} constrains $\kappa$, for $m_Z
- \Gamma_{Z} \leqslant m_{Z'} \leqslant m_Z + \Gamma_{Z}$ where we
consider Eq.~\eqref{eq:Gamma-inv-exp} to hold true as a
phenomenological approximation. As shown in
Fig.~\ref{fig:kappa-mZp-Nnu}, the scaling factor $\kappa$ is
constrained to be
\begin{align*}
    & 0 \leqslant \kappa \leqslant 0.0356 & \textrm{ at $1\sigma$
    level for } m_{Z'} = m_Z - \Gamma_Z,\\*%
    & 0 \leqslant \kappa \leqslant 0.0346 & \textrm{ at $1\sigma$
    level for } m_{Z'} = m_Z + \Gamma_Z.
\end{align*}

We emphasize that we are assuming the existence of a
quasi-degenerate $Z'$ boson which is hidden or unresolved in the
present experimental data\footnote{The width of a
quasi-degenerate $Z'$ is certainly narrower than that of the $Z$
boson, as can be seen from Eq.~\eqref{eq:Gamma_Z_Zp_relation} with
$\kappa < 1$. It is plausible that in the binned experimental data the
size of the bin might be larger than or comparable to the width of
$Z'$, such that the associated dip and peak contribution from $Z'$
very slightly alter the $Z$ line shape, thereby evading a direct
detection so far.}. Thus, it contributes to
experimentally inferred invisible width, and the reported value of the
number of light neutrinos $N_{\nu}$ can constrain the coupling of $Z'$
to neutrinos, in general. This constraint applies to all the couplings
of $Z'$ to SM fermions when working under our assumptions of
Eqs.~\eqref{eq:kappa_V_kappa_A} and \eqref{eq:kappa}.

We reiterate that in the current article we consider NP contribution
\textit{solely} arising from an extra physical $Z'$ boson `practically
degenerate' with the SM $Z$ boson, i.e.\ the couplings of SM $Z$
boson to fermions is not affected by having an extra $Z'$ in the
picture. Of course, we can instead consider a new physics which comes
directly into the SM $Z$ boson, and affect the $Z \to \nu \bar\nu$
branching ratio and thus $N_{\nu}$, see e.g.\ \cite{Kim:2022xjg} for
model-independent constraints on such NP possibility.

\subsection{Cross-section and forward-backward asymmetry}\label{subsec:pheno-cross-section}

To see the effects of a quasi-degenerate $Z'$ in a more clear manner,
we need to consider both its on-shell production and decay. Regarding
this, one might consider either $e^+ \, e^-$ collisions as done
previously at LEP, or $p \, p$ collisions as currently done at LHC, or
even $\mu^+ \, \mu^-$ collisions as envisioned in future muon
colliders \cite{InternationalMuonCollider:2025sys}. As $e^+ e^-$
colliders have played a vital role in the discovery of the $Z$ boson
and due to their much cleaner backgrounds, we choose to consider the
production and decay of $Z'$ in such a collider. Phenomenologically,
the processes of our interest are the $s$-channel events of $e^- e^+
\to f \bar{f}$ with $f \neq e^-$ (to avoid consideration of Bhabha
scattering which has additional $t$-channel contributions).

\begin{figure}[hbtp]
\centering%
\includegraphics[scale=1]{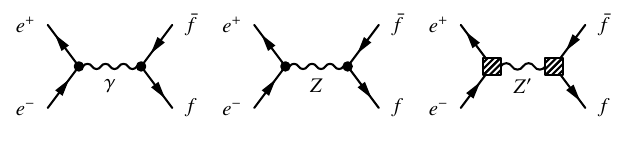}%
\caption{Tree-level Feynman diagrams for the process $e^-e^+ \to f
\bar{f}$ with $f \neq e^-$. The square blobs on the vertices involving
$Z'$ denote the vertex factors from Eq.~\eqref{eq:gen-Lag-Zpff-v2}. In
this article we do not consider any beyond tree-level contributions,
such as those mediated via various quarkonium possibilities.}%
\label{fig:FD}
\end{figure}

In our case, the process $e^- e^+ \to f \bar{f}$ with $f \neq e^-$
takes place via $s$-channel $\gamma$, $Z$ and $Z'$ mediated Feynman
diagrams at tree-level, see Fig.~\ref{fig:FD}. We confine ourselves to
tree-level contributions alone as they are sufficient to show the
signatures of the presence of $Z'$ on the observables. Neglecting
fermion masses with respect to the mass of the $Z$ boson, we find that
the differential cross-section for $e^- e^+ \to f \bar{f}$ can be
written as,
\begin{equation}\label{eq:diff-cross-section}
\frac{\dd\sigma_{ff}(s,\kappa)}{\dd\cos\theta} = \frac{\pi}{2\,s}
N_c^f \left(\sigma_E^f(s,\kappa) \, \big(1+\cos^2\theta\big) +
\sigma_O^f(s,\kappa) \, \cos\theta\right),
\end{equation}
where $N_c^f$ takes care of the color possibilities for fermion $f$,
$\sqrt{s}$ is the total energy in the center-of-momentum frame of $e^-
e^+$, and the angle $\theta$ is defined as the angle between
directions of flights of $e^-$ and $f$ in the center-of-momentum frame
of $e^- e^+$. The differential cross-section of
Eq.~\eqref{eq:diff-cross-section} has a part proportional to
$\cos\theta$ which we have denoted by $\sigma_O^f(s,\kappa)$ and a
part proportional to $1 +\cos^2\theta$ which we have denoted by
$\sigma_E^f(s,\kappa)$. Both $\sigma_E^f (s)$ and $\sigma_O^f (s)$
contain contributions from the individual diagrams of
Fig.~\ref{fig:FD} as well as their interference terms,
\begin{align}
\sigma_E^f(s,\kappa) &= \alpha_{\textrm{EM}}^2 q_f^2 - 8 \,
\alpha_{\textrm{EM}} \, q_f \, g_V^e \, g_V^f \, \left( \Re(r_{Z}(s))
+ \kappa^2 \, \Re(r_{Z'}(s,\kappa)) \right) \nonumber\\%
&\quad + 16 \, \left( \big(g_V^e\big)^2 + \big(g_A^e\big)^2 \right)
\left( \big(g_V^f\big)^2 + \big(g_A^f\big)^2 \right) \,
\modulus{r_{Z}(s) + \kappa^2 \, r_{Z'}(s,\kappa)}^2,
\label{eq:sigma_E}\\%
\sigma_O^f(s,\kappa) &= -16 \, \alpha_{\textrm{EM}} \, q_f \, g_A^e \,
g_A^f \, \left( \Re(r_{Z}(s)) + \kappa^2 \, \Re(r_{Z'}(s,\kappa))
\right) \nonumber\\*%
&\quad + 128 \, g_V^e \, g_A^e \, g_V^f \, g_A^f \, \modulus{r_{Z}(s)
+ \kappa^2 \, r_{Z'}(s,\kappa)}^2, \label{eq:sigma_O}
\end{align}
with $\alpha_{\textrm{EM}} = e^2/(4\pi)$ being the electromagnetic
fine-structure constant, $q_f$ being the electric charge of the
fermion $f$, and $r_{Z}(s)$, $r_{Z'}(s)$ contain the $s$-dependent
Breit-Wigner resonance forms for $Z$ and $Z'$
respectively\footnote{The photon pole at $s=0$ is avoided by imposing
the fact that the minimum value of $s$ for $e^-e^+ \to f\bar{f}$ is
$s_{\textrm{min}}=4 m_f^2 >0$ with $m_f$ being the mass of fermion
$f$. Also note that here we are only analyzing the $Z'$
phenomenology at tree-level. The correction to $Z$ and $Z'$
propagators arising from loop-induced off-diagonal self-energy terms
\cite{LoChiatto:2024guj} ought to be considered in a full
loop-corrected calculation.},
\begin{subequations}\label{eq:rZ_rZp}
\begin{align}
r_{Z}(s) &= \left( \frac{g_{Z}^2}{64\pi} \right) \, \left( \frac{s}{s
- m_{Z}^2 + i \, s \, \Gamma_{Z}/m_Z} \right) = \left( \frac{G_F \,
m_Z^2}{8\pi\sqrt{2}} \right) \, \left( \frac{s}{s - m_{Z}^2 + i \, s
\, \Gamma_{Z}/m_{Z} } \right),\label{eq:rZ}\\%
r_{Z'}(s,\kappa) &= \left( \frac{g_{Z}^2}{64\pi} \right) \, \left(
\frac{s}{s - m_{Z'}^2 + i \, s \, \Gamma_{Z'}/m_{Z'}} \right) = \left(
\frac{G_F \, m_Z^2}{8\pi\sqrt{2}} \right) \, \left( \frac{s}{s -
m_{Z'}^2 + i \, \kappa^2 \, s \, \Gamma_{Z}/m_{Z}}
\right),\label{eq:rZp}
\end{align}
\end{subequations}
where we have used Eq.~\eqref{eq:Gamma_Z_Zp_relation} to make the
substitution $\Gamma_{Z'}/m_{Z'} = \kappa^2 \, \Gamma_{Z}/m_{Z}$.

The total cross-section for $e^- e^+ \to f \, \bar{f}$ is then given by,
\begin{equation}\label{eq:cross-section-full}
\sigma_{ff} (s,\kappa) \equiv \int_{-1}^{+1}
\left(\frac{\dd\sigma_{ff} (s,\kappa) }{\dd\cos\theta}\right) \,
\dd\cos\theta = \frac{4\pi}{3s} \, N_c^f \, \sigma_E^f(s,\kappa),
\end{equation}
and the distribution asymmetry between forward $\left( 0 \leqslant
\cos\theta \leqslant 1 \right)$ and backward $\left( -1 \leqslant
\cos\theta \leqslant 0 \right)$ regions, known as the forward-backward
asymmetry is given by,
\begin{equation}\label{eq:AFB-full}
A_{FB}^{ff} (s,\kappa) \equiv \dfrac{\displaystyle \int_{0}^{+1}
\left(\dfrac{\dd\sigma_{ff} (s,\kappa) }{\dd\cos\theta}\right) \,
\dd\cos\theta - \int_{-1}^{0} \left(\dfrac{\dd\sigma_{ff}
(s,\kappa)}{\dd\cos\theta}\right) \, \dd\cos\theta}{\displaystyle
\int_{-1}^{+1} \left(\dfrac{\dd\sigma_{ff}
(s,\kappa)}{\dd\cos\theta}\right) \, \dd\cos\theta}= \frac{3}{8} \,
\frac{\sigma_O^f (s,\kappa)}{\sigma_E^f (s,\kappa)}.
\end{equation}
Note that in Eqs.~\eqref{eq:cross-section-full} and
\eqref{eq:AFB-full} the angular integrations are done assuming an
ideal detector with full $4\pi$ solid angle coverage around the point
of collision. In reality, detectors do not cover the full $4\pi$ solid
angle and the corresponding angular coverage should appropriately get
reflected in the integration ranges. For a given arbitrary detector
whose detector acceptance is a practically uniform function of
$\cos\theta$, let us assume that $-x \leqslant \cos\theta \leqslant
+x$ with $0 < x <1$ and without any detector malfunctions or
anomalous effects in the said ranges. In such a case, the total
cross-section and the forward-backward asymmetry get modified as
follows,
\begin{subequations}
\begin{align}
\sigma_{ff}^{\textrm{cut}} (s,\kappa) &\equiv \int_{-x}^{+x}
\left(\frac{\dd\sigma_{ff} (s,\kappa) }{\dd\cos\theta}\right) \,
\dd\cos\theta = \left( \frac{x\left(x^2+3\right)}{4} \right) \,
\sigma_{ff}(s,\kappa), \label{eq:cross-section-cut}\\%
A_{FB}^{ff,\textrm{ cut}} (s,\kappa) &\equiv \dfrac{\displaystyle
\int_{0}^{+x} \left(\dfrac{\dd\sigma_{ff} (s,\kappa)
}{\dd\cos\theta}\right) \, \dd\cos\theta - \int_{-x}^{0}
\left(\dfrac{\dd\sigma_{ff} (s,\kappa)}{\dd\cos\theta}\right) \,
\dd\cos\theta}{\displaystyle \int_{-x}^{+x}
\left(\dfrac{\dd\sigma_{ff} (s,\kappa)}{\dd\cos\theta}\right) \,
\dd\cos\theta}= \left( \frac{4\,x}{x^2+3} \right) \, A_{FB}^{ff}
(s,\kappa). \label{eq:AFB-cut}
\end{align}
\end{subequations}

Note that both the cross-section $\sigma_{ff}(s,\kappa)$ and
forward-backward asymmetry $A_{FB}^{ff}(s,\kappa)$ as given in
Eqs.~\eqref{eq:cross-section-full} and \eqref{eq:AFB-full}, as well as
$\sigma_{ff}^{\textrm{cut}}(s,\kappa)$ and
$A_{FB}^{ff,\textrm{cut}}(s,\kappa)$ as given in
Eqs.~\eqref{eq:cross-section-cut} and \eqref{eq:AFB-cut} respectively,
are analytic functions of $s$, since we have used Breit-Wigner
resonance factors for the $Z$ and $Z'$ contributions.

\subsection{Difference-over-sum ratios as observables}

We note that imposition of angular cuts only introduce overall
multiplicative factors, essentially scaling the un-cut cross-section
and forward-backward asymmetry. Nevertheless, it is important to note
that for $\kappa \neq 0$, the quantities $\sigma_{ff}^{\textrm{cut}}
(s,\kappa)$ and $A_{FB}^{ff,\textrm{ cut}} (s,\kappa)$, or
$\sigma_{ff} (s,\kappa)$ and $A_{FB}^{ff} (s,\kappa)$, contain both SM
as well as NP (i.e.\ $Z'$) contributions. If we put $\kappa=0$, then
we get the SM contribution alone. It is interesting to note that we
can define the following two dimensionless observables defined as
difference-over-sum (DoS) ratios which are especially insensitive to
the specific angular cuts of any detector,
\begin{subequations}\label{eq:DoS}
\begin{align}
\mathcal{A}_{\sigma}^{ff} (s,\kappa) &=
\frac{\sigma_{ff}^{\textrm{cut}} (s,\kappa) -
\sigma_{ff}^{\textrm{cut}} (s,0)}{\sigma_{ff}^{\textrm{cut}}
(s,\kappa) + \sigma_{ff}^{\textrm{cut}} (s,0)} = \frac{\sigma_{ff}
(s,\kappa) - \sigma_{ff} (s,0)}{\sigma_{ff} (s,\kappa) + \sigma_{ff}
(s,0)}, \label{eq:DoS_sigma}\\%
\mathcal{A}_{A_{FB}}^{ff} (s,\kappa) &= \frac{A_{FB}^{ff,\textrm{
cut}} (s,\kappa) - A_{FB}^{ff,\textrm{ cut}}
(s,0)}{A_{FB}^{ff,\textrm{ cut}} (s,\kappa) + A_{FB}^{ff,\textrm{
cut}} (s,0)} = \frac{A_{FB}^{ff} (s,\kappa) - A_{FB}^{ff}
(s,0)}{A_{FB}^{ff} (s,\kappa) + A_{FB}^{ff} (s,0)}. \label{eq:DoS_AFB}
\end{align}
\end{subequations}
By construction, $-1 \leqslant \mathcal{A}_{\sigma}^{ff} (s,\kappa)
\leqslant +1$ as the cross-section is always non-zero in the mass
region of our interest. But $\mathcal{A}_{A_{FB}}^{ff} (s,\kappa)$ can
shoot beyond the range $[-1,1]$ as the forward-backward asymmetry can
vanish for some value of $s$ and $\kappa$. Moreover, one can encounter
singularity in $\mathcal{A}_{A_{FB}}^{ff} (s,\kappa)$ when $\big(
A_{FB}^{ff,\textrm{ cut}} (s,\kappa) + A_{FB}^{ff,\textrm{ cut}} (s,0)
\big) \to 0$ or $\big(A_{FB}^{ff} (s,\kappa) + A_{FB}^{ff} (s,0)\big)
\to 0$.  Nevertheless, the two quantities in Eq.~\eqref{eq:DoS}
``contrast'' the presence of NP signal with respect to the SM
expectation.

It is important to note that probing the underlying NP parameter space
using a set of complimentary but independent observables is a more
robust and credible phenomenological approach than relying on one
particular observable alone. From this perspective, all the
observables mentioned thus far can be used together for probing the
quasi-degenerate $Z'$ possibility. 
Most importantly, although $N_\nu$ serves as an independent quantity
extracted from precision $Z$-pole studies, the other observables
$\sigma_{\mu\mu}(s,\kappa)$, $A_{FB}^{\mu\mu}(s,\kappa)$,
$\mathcal{A}_{\sigma}^{\mu\mu} (s,\kappa)$ and
$\mathcal{A}_{A_{FB}}^{\mu\mu}(s,\kappa)$ can be measured above, below
and near the $Z$ and $Z'$ poles\footnote{Such study of the observables
away from $Z$ pole would be both meaningful and helpful in context of
future colliders which are proposed to operate at higher
center-of-momentum energies than present and past experiments, e.g.\
in international muon collider
\cite{InternationalMuonCollider:2025sys}. Additionally, there
are also proposals to study the $Z$-pole region with more data in
future colliders such as the FCC-ee~\cite{Toffolin:2026rbd,
FCC:2025lpp, FCC:2025uan, FCC:2018byv, FCC:2018evy, Blondel:2018mad},
CEPC~\cite{CEPCStudyGroup:2018rmc, CEPCStudyGroup:2018ghi, LouCEPC:2019nrp, 
CEPCStudyGroup:2023quu} and ILC~\cite{ILC:2007bjz, ILC:2013jhg,
Bambade:2019fyw} in addition to measurements away from $Z$-pole.}.

\section{Numerical study and discussions}\label{sec:numerical-discussion}

\begin{figure}[hbtp]
\centering%
\subfloat[]{\includegraphics{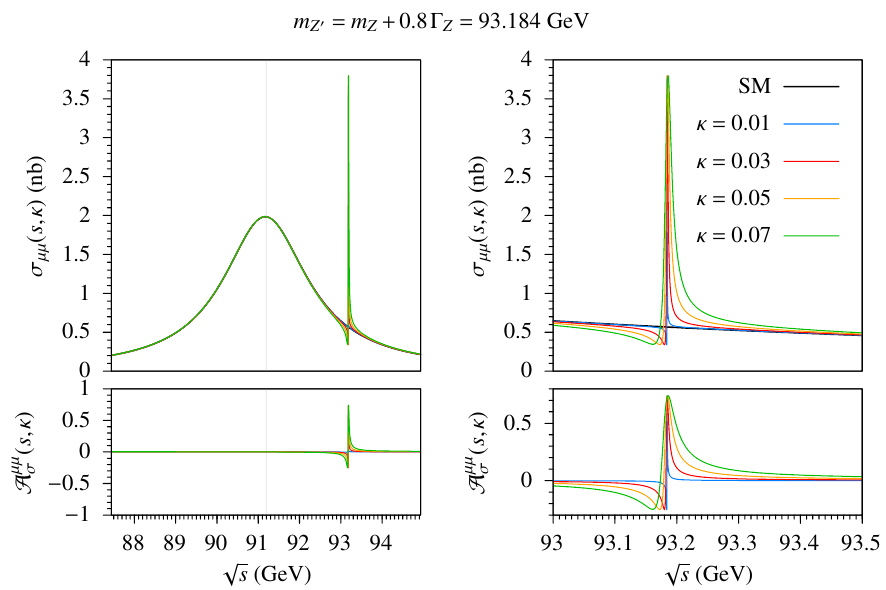}}\\%
\subfloat[]{\includegraphics{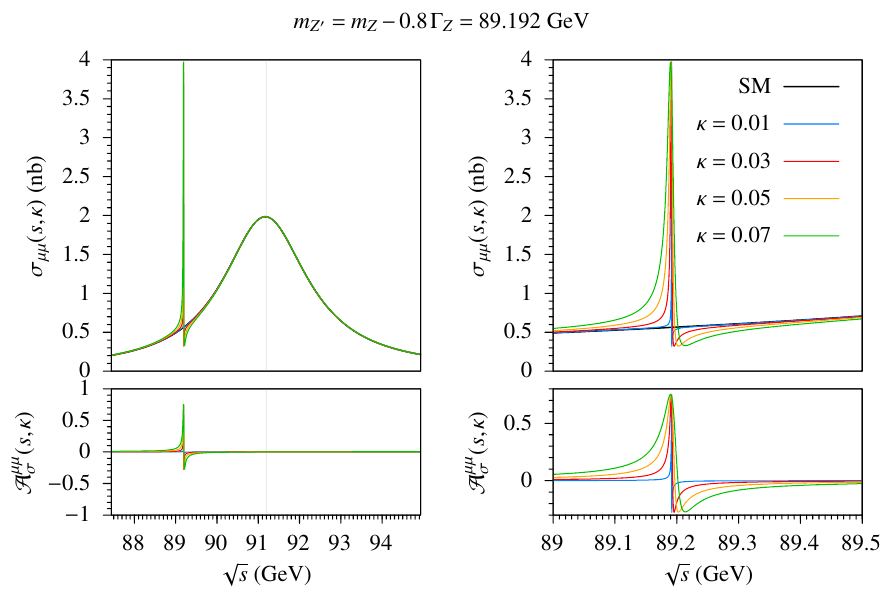}}%
\caption{Illustration of how the cross-section
$\sigma_{\mu\mu}(s,\kappa)$ as well as the corresponding DoS ratio
$\mathcal{A}_{\sigma}^{\mu\mu} (s,\kappa)$ get affected in presence of
a quasi-degenerate $Z'$ with mass $m_{Z'} = m_Z \pm 0.8 \,
\Gamma_{Z}$. We have considered values of $\kappa \leqslant 0.07$
respecting the constraint from $N_\nu$ as shown in
Fig.~\ref{fig:kappa-mZp-Nnu}. The thin vertical light gray line
corresponds to $\sqrt{s} = m_Z$.}%
\label{fig:cross-section-full_DoS}
\end{figure}

\begin{figure}[hbtp]
\centering%
\subfloat[]{\includegraphics{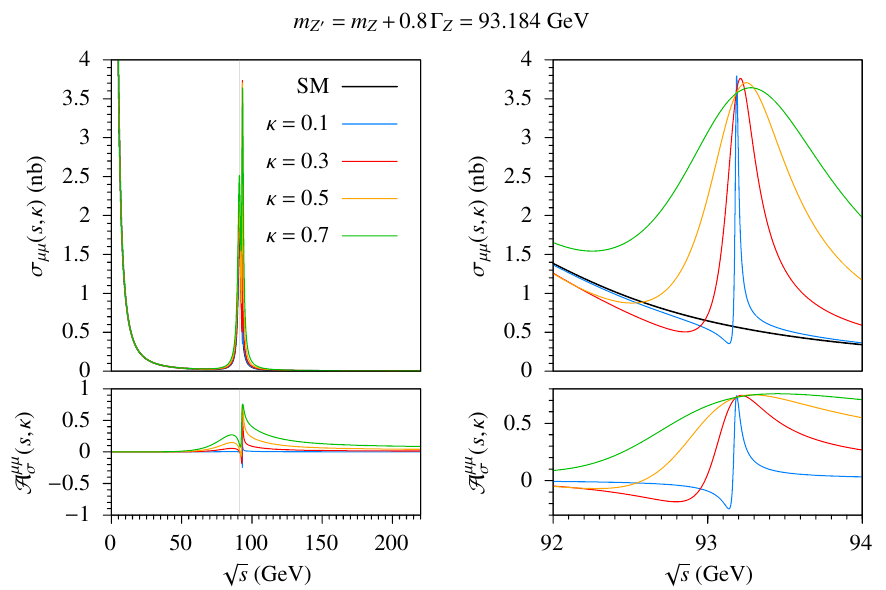}}\\%
\subfloat[]{\includegraphics{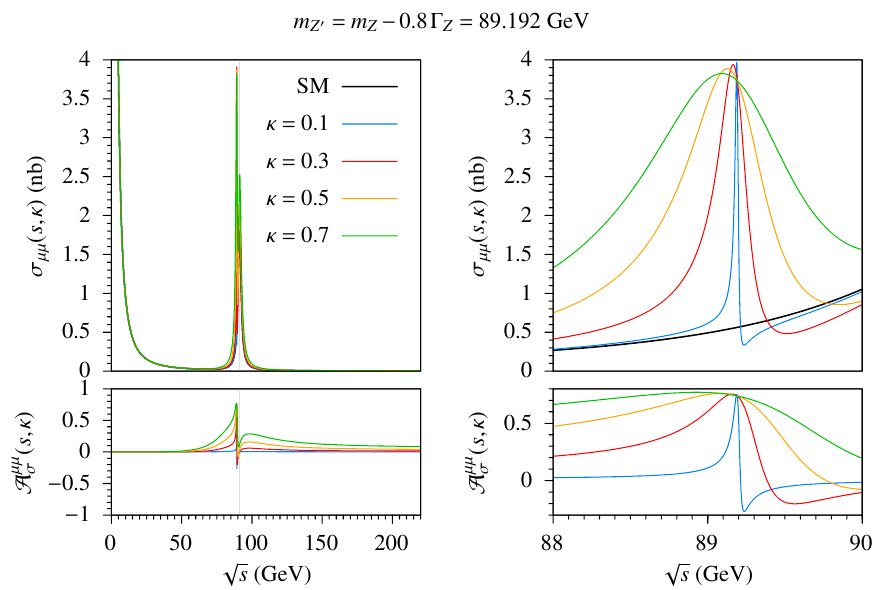}}%
\caption{Similar illustration as given in
Fig.~\ref{fig:cross-section-full_DoS} but here we consider the
possibility of $\kappa > 0.07$ by disregarding the constraint from
$N_\nu$ as shown in Fig.~\ref{fig:kappa-mZp-Nnu}. The thin vertical
light gray line corresponds to $\sqrt{s} = m_Z$. Notice that although
the effect on $\sigma_{\mu\mu}(s,\kappa)$ is visually localized in the
neighbourhood of the $Z'$ resonance. On the other hand, the
corresponding DoS ratio $\mathcal{A}_{\sigma}^{\mu\mu}(s,\kappa)$ is
visibly different even away from the $Z'$ resonances. No quarkonium
states are considered here.}%
\label{fig:cross-section-full_DoS_large_kappa}
\end{figure}

\begin{figure}[hbtp]
\centering%
\subfloat[]{\includegraphics{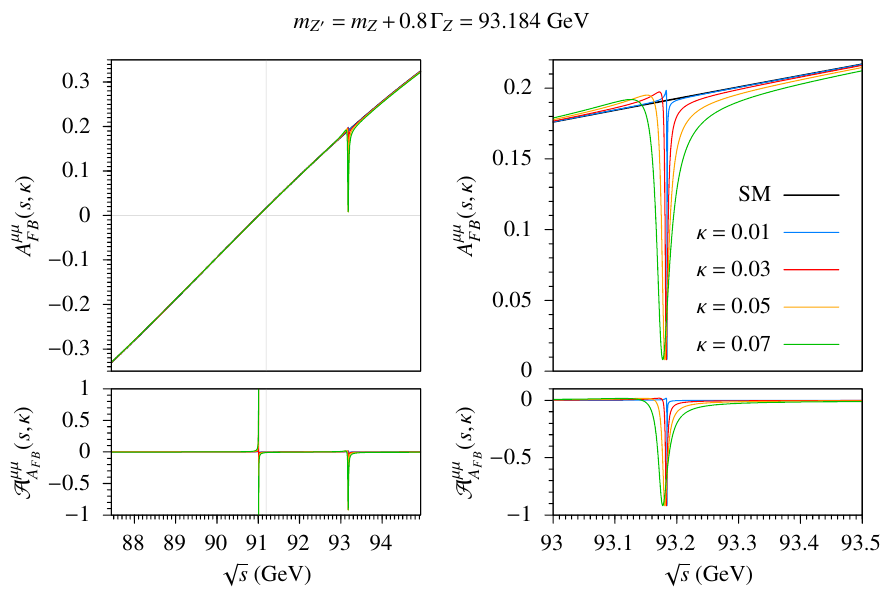}}\\%
\subfloat[]{\includegraphics{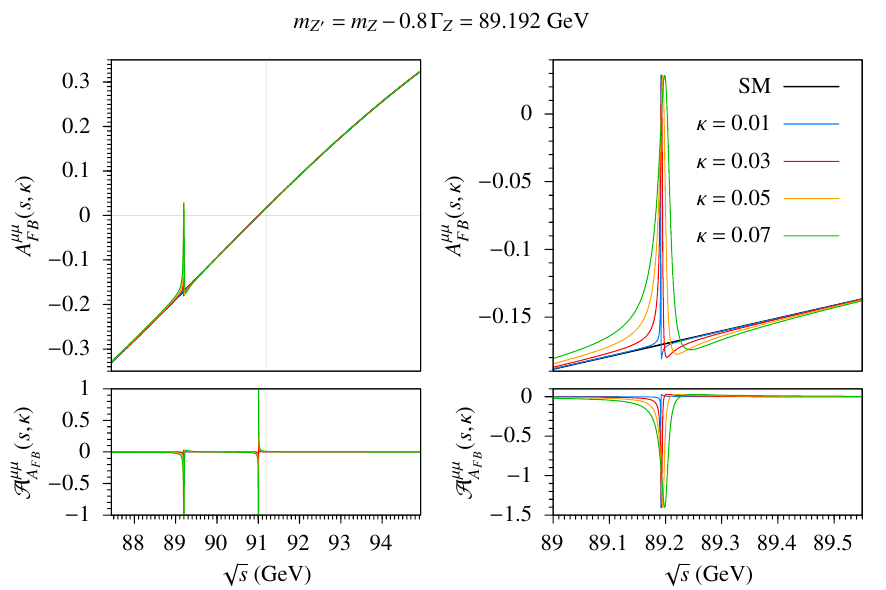}}%
\caption{Illustration of how the forward-backward asymmetry
$A_{FB}^{\mu\mu}(s,\kappa)$ as well as the corresponding DoS ratio
$\mathcal{A}_{A_{FB}}^{\mu\mu} (s,\kappa)$ get affected in presence of
a quasi-degenerate $Z'$ with mass $m_{Z'} = m_Z \pm 0.8 \,
\Gamma_{Z}$. We have considered values of $\kappa \leqslant 0.07$
respecting the constraint from $N_\nu$ as shown in
Fig.~\ref{fig:kappa-mZp-Nnu}. The thin vertical light gray line
corresponds to $\sqrt{s} = m_Z$. Note that we encounter singularities
in $\mathcal{A}_{A_{FB}}^{\mu\mu} (s,\kappa)$ in the limit $\left(
A_{FB}^{\mu\mu}(s,\kappa) + A_{FB}^{\mu\mu}(s,0) \right) \to 0$, which
in this case occurs at $\sqrt{s} \approx 91$~GeV. The negative spikes
that one observes around $\sqrt{s} = m_{Z'}$ do not correspond to
singularities in $\mathcal{A}_{A_{FB}}^{\mu\mu} (s,\kappa)$ but in
plot (b) the value of $\mathcal{A}_{A_{FB}}^{\mu\mu} (s,\kappa)$ goes
below $-1$ showing that $\mathcal{A}_{A_{FB}}^{\mu\mu} (s,\kappa)$ is
not bounded in the region $[-1,1]$.}%
\label{fig:AFB-full_DoS}
\end{figure}

\begin{figure}[hbtp]
\centering%
\subfloat[]{\includegraphics{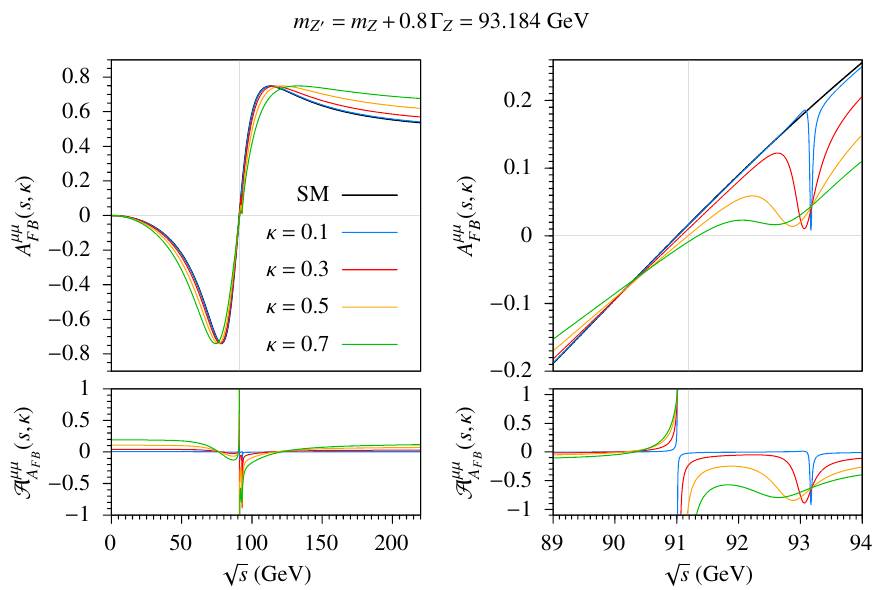}}\\%
\subfloat[]{\includegraphics{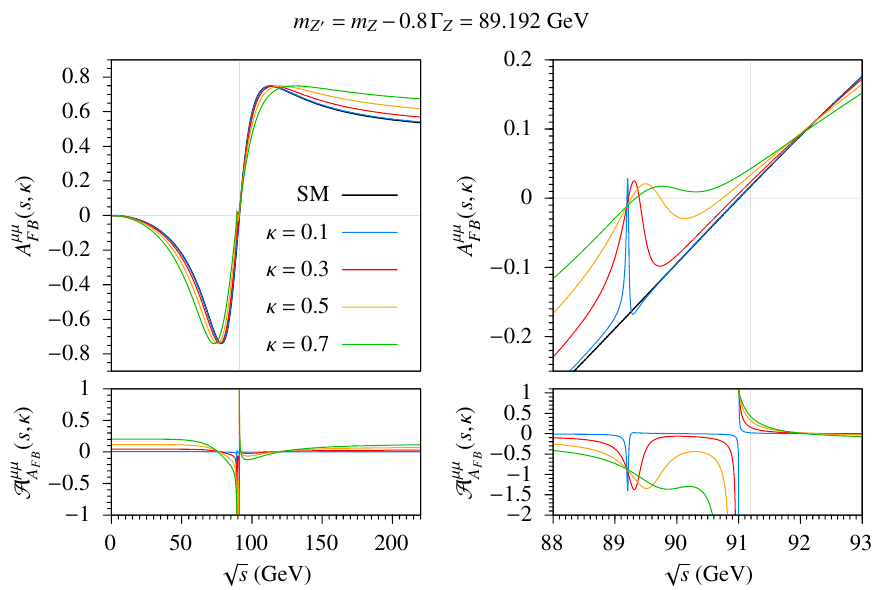}}%
\caption{Similar illustration as given in Fig.~\ref{fig:AFB-full_DoS}
but here we consider the possibility of $\kappa > 0.07$ by
disregarding the constraint from $N_\nu$ as shown in
Fig.~\ref{fig:kappa-mZp-Nnu}. The thin vertical light gray line
corresponds to $\sqrt{s} = m_Z$. Once again note that we encounter
singularities in $\mathcal{A}_{A_{FB}}^{\mu\mu} (s,\kappa)$ in the
limit $\left( A_{FB}^{\mu\mu}(s,\kappa) + A_{FB}^{\mu\mu}(s,0) \right)
\to 0$. Most importantly, the visibility of the effect of an
additional $Z'$ is not confined to the neighbourhood of $m_{Z'}$
unlike Figs.~\ref{fig:cross-section-full_DoS} and
\ref{fig:AFB-full_DoS}. No quarkonium states are considered here.}%
\label{fig:AFB-full_DoS_large_kappa}
\end{figure}

For a numerical study of the observables mentioned before, we consider
the specific scattering process $e^+ \, e^- \to \mu^+ \, \mu^-$. We
consider two different quasi-degenerate $Z'$ possibilities
corresponding to
\begin{equation*}
    m_{Z'} = m_Z \pm 0.8\, \Gamma_{Z},
\end{equation*}
for illustration and further discussion. The effects of such a $Z'$ on
the cross-section, forward-backward asymmetry as well as their
corresponding DoS ratios are shown in
Figs.~\ref{fig:cross-section-full_DoS},
\ref{fig:cross-section-full_DoS_large_kappa}, \ref{fig:AFB-full_DoS}
and \ref{fig:AFB-full_DoS_large_kappa}. In
Figs.~\ref{fig:cross-section-full_DoS} and \ref{fig:AFB-full_DoS} we
consider a few representative values of $\kappa$ satisfying $\kappa
\leqslant 0.07$ by taking into account the $3\sigma$ allowed
constraint from $N_\nu$ as shown in Fig.~\ref{fig:kappa-mZp-Nnu}. The
effects of the $Z'$ on the concerned distributions in
Figs.~\ref{fig:cross-section-full_DoS} and \ref{fig:AFB-full_DoS} are
localized in the neighbourhood of $\sqrt{s} = m_{Z'}$, i.e.\ close to
the $Z'$ pole. It is plausible that for a smaller value of $\kappa$,
the effects of $Z'$ might so localized that those become noticeable
only when the experimental resolution of $\sqrt{s}$ is significantly
improved. On the other hand, in
Figs.~\ref{fig:cross-section-full_DoS_large_kappa} and
\ref{fig:AFB-full_DoS_large_kappa} we consider larger values of
$\kappa$ (satisfying $0.07 < \kappa < 1$) with an aim to highlight the
fact that such large values of $\kappa$ could be constrained
independently\footnote{In a very contrived but different case of
a quasi-degenerate $Z'$ which is neutrino-phobic, i.e.\ $\kappa_V^\nu
= 0 = \kappa_A^\nu$ while $\kappa_V^f = \kappa_A^f = \kappa \neq 0$ for all
other fermions $f$, one can entirely bypass the constraint from
$N_\nu$. Then, the scaling factor $\kappa$ for such a $Z'$ could be
probed and/or constrained using cross-section and $A_{FB}$ as well as
the corresponding DoS ratios.} from observables other than the
$N_\nu$. Figs.~\ref{fig:cross-section-full_DoS_large_kappa} and
\ref{fig:AFB-full_DoS_large_kappa} clearly show that the
forward-backward asymmetry and the DoS ratios corresponding to both
the cross-section and the forward-backward asymmetry are capable of
constraining even a quasi-degenerate $Z'$ by looking at experimental
measurements farther from $\sqrt{s} = m_{Z'}$ on both directions. Most
importantly, the trend in Fig.~\ref{fig:AFB-full_DoS_large_kappa}
implies that precision studies of $A_{FB}$ and its corresponding DoS
ratio $\mathcal{A}_{A_{FB}}^{\mu\mu} (s,\kappa)$ can possibly
constrain the quasi-degenerate $Z'$ possibilities even farther away
from $Z$ pole.

\begin{table}[hbtp]
\centering%
\begin{tabular}{c|c|c|c|c|c|c|c|c}\toprule%
\multirow{2}{*}{$m_{Z'}$} & \multicolumn{8}{|c}{$\kappa$} \\ \cline{2-9}%
& $0.01$ & $0.03$ & $0.05$ & $0.07$ & $0.1$ & $0.3$ & $0.5$ & $0.7$\\ \midrule%
$m_Z - 0.8\,\Gamma_Z$ & $91.0069$ & $91.0066$ & $91.0059$ & $91.0049$ & $91.0028$ & $90.9671$ & $90.8895$ & $90.7811$\\ \hline%
$m_Z + 0.8\,\Gamma_Z$ & $91.0070$ & $91.0073$ & $91.0078$ & $91.0087$ & $91.0104$ & $91.0372$ & $91.0856$ & $91.1354$\\ \bottomrule%
\end{tabular}
\caption{The singulrities of $\mathcal{A}_{A_{FB}}^{\mu\mu}
(s,\kappa)$ as apparent in Figs.~\ref{fig:AFB-full_DoS} and
\ref{fig:AFB-full_DoS_large_kappa} happens at the $\sqrt{s}$
values as mentioned (in units of GeV) in this table, for the
different $m_{Z'}$ and $\kappa$ values we have considered.}%
\label{tab:singularities}
\end{table}

The singularities in $\mathcal{A}_{A_{FB}}^{\mu\mu} (s,\kappa)$ in the
limit $\left( A_{FB}^{\mu\mu}(s,\kappa) + A_{FB}^{\mu\mu}(s,0) \right)
\to 0$ are quite clearly visible in the left-hand side panels of
Figs.~\ref{fig:AFB-full_DoS} and \ref{fig:AFB-full_DoS_large_kappa}.
These mathematical singularities are an artifact of the
normalization used in Eq.~\eqref{eq:DoS_AFB} and are not associated
with presence of any physical resonances, and they appear at
$\sqrt{s} \approx 91$~GeV for almost all values of $\kappa$ and
$m_{Z'}$ we have considered in Figs.~\ref{fig:AFB-full_DoS} and
\ref{fig:AFB-full_DoS_large_kappa}, as can also be seen from the
values tabulated in Table~\ref{tab:singularities}. The negative spikes
one finds near $\sqrt{s}=m_{Z'}$ in the left-hand side panels of
Figs.~\ref{fig:AFB-full_DoS} and \ref{fig:AFB-full_DoS_large_kappa} do
not correspond to any singularities, as can be seen from the zoomed-in
plots on the right-hand panels of the corresponding figures. It can be
easily seen from Figs.~\ref{fig:cross-section-full_DoS},
\ref{fig:AFB-full_DoS}, \ref{fig:cross-section-full_DoS_large_kappa}
and \ref{fig:AFB-full_DoS_large_kappa} (especially by looking at the
right-hand panels of those figures) that the width of the peak or dip
near $\sqrt{s}=m_{Z'}$ increases with larger value of $\kappa$ when
$m_{Z'}$ is kept fixed, $\Gamma_{Z'} \propto \kappa^2$ in accordance
with the expected from Eq.~\eqref{eq:Gamma_Z_Zp_relation}. This
$\kappa$ dependent $\Gamma_{Z'}$ enters the Breit-Wigner form of the
resonance via Eq.~\eqref{eq:rZp}.

Cross-section $\sigma_{ff}(s,\kappa)$ and forward-backward asymmetry
$A_{FB}^{ff}(s,\kappa)$ are well-known experimental observables in the
literature. However, as mentioned before, they suffer from the limited
angular coverage allowed by the detector  of every experiment that
studies the underlying process. It is important to note the DoS ratios
of Eq.~\eqref{eq:DoS} can help one overcome that limitation to compare
different experimental results more uniformly.

More importantly, the DoS ratios can be directly implemented for every
experimental setup by observing that
\begin{enumerate}
\item $\sigma_{ff}^{\textrm{cut}} (s,\kappa)$ and $A_{FB}^{ff,\textrm{
cut}} (s,\kappa)$ are essentially the experimentally observed
cross-section $\sigma_{ff}^{\textrm{Expt}}(s)$ and forward-backward
asymmetry $A_{FB}^{ff,\textrm{ Expt}}(s)$ respectively, and

\item $\sigma_{ff}^{\textrm{cut}} (s,0)$ and $A_{FB}^{ff,\textrm{
cut}} (s,0)$ essentially correspond to the SM theory predictions for
cross-section $\sigma_{ff}^{\textrm{SM Th}}(s)$ and forward-backward
asymmetry $A_{FB}^{ff,\textrm{ SM Th}}(s)$ respectively, taking into
account the specific experimental constraints.
\end{enumerate}
This implies that Eq.~\eqref{eq:DoS} can be rewritten in the following
form,
\begin{equation}\label{eq:DoS-practical}
\mathcal{A}_{\sigma}^{ff} (s) = \frac{\sigma_{ff}^{\textrm{Expt}}(s) -
\sigma_{ff}^{\textrm{SM Th}}(s)}{\sigma_{ff}^{\textrm{Expt}}(s) +
\sigma_{ff}^{\textrm{SM Th}}(s)}, \qquad%
\mathcal{A}_{A_{FB}}^{ff} (s) = \frac{A_{FB}^{ff,\textrm{ Expt}}(s) -
A_{FB}^{ff,\textrm{ SM Th}}(s)}{A_{FB}^{ff,\textrm{ Expt}}(s) +
A_{FB}^{ff,\textrm{ SM Th}}(s)}.
\end{equation}
We are of the opinion that it would be useful if in future these
observables can also be studied along with the usual cross-sections
and forward-backward asymmetries. We also clarify that taking all past
experimental results on $e^+ e^- \to \mu^+ \mu^-$ and doing a
phenomenological least-square analysis to constrain $m_{Z'}$ and
$\kappa$ currently lies beyond our expertise. It must be done
diligently by those well versed in the experimental matters taking
care of all detector aspects, spread of beam energy etc.\ as well as
higher-order theory contributions by going beyond our tree-level
considerations done here.

Even if such careful studies in future could reveal stringent
experimental constraints on such quasi-degenerate $Z'$, those can be
evaded if one were to assume that the $Z'$ boson does not have family
universal couplings, such as say a muon-philic (or neutrino-philic)
$Z'$ which couples to only muons (or neutrinos) and not other fermions
such that it could not have been produced in $e^- e^+$ colliders, nor
could it be produced in $pp$ colliders. The muon-philic $Z'$ could
certainly be produced on-shell in a future $\mu^- \mu^+$ collider of
appropriate center-of-momentum energy, and one could analyze the
cross-section, forward-backward asymmetry as well as the corresponding
DoS ratios for the process $\mu^- \mu^+ \to \mu^- \mu^+$ mediated via
both $s$ and $t$ channel exchanges of photon, $Z$ and $Z'$. A
neutrino-philic $Z'$, on the other hand, would be impossible to
produce and detect in any detector, as it couples exclusively to
neutrinos and it remains to be explored whether such a $Z'$ could even
be constrained indirectly. We reserve consideration of such exotic
quasi-degenerate $Z'$'s for future explorations if such a need arises.

\section{Conclusion}\label{sec:conclusion}

The curious possibility of a $Z'$ boson around the SM $Z$-pole
with $\modulus{m_Z - m_{Z'}} \lesssim \Gamma_Z$ which has stayed
unresolved or hidden in present experimental data is a very
interesting possibility that can be probed and constrained in future
collider searches. Instead of constructing a UV-complete model with
such a possible $Z'$, we dwell on a purely model-independent approach
to study the phenomenology of such a quasi-degenerate $Z'$ at
tree-level. Lorentz invariance and renormalizability suggest SM
$Z$-like couplings to fermions for such a $Z'$. For a simplified
phenomenological analysis we impose two simplifying assumptions of
family universality and uniform scaling of vector and axial-vector
couplings by a scaling factor $\kappa$. Presence of such an
experimentally unresolved quasi-degenerate $Z'$ can potentially
contribute to the measured invisible width at the $Z$-pole. We find
that the measured value of the number of light neutrinos $N_{\nu}$
constrains the coupling of $Z'$ to SM fermions, specifically the
scaling factor $\kappa$ is constrained to be $0 \leqslant \kappa
\lesssim 0.035$ at $1\sigma$ level. We also show how the well-known
observables of cross-section and forward-backward asymmetry ($A_{FB}$)
for $e^- e^+ \to f \bar{f}$ scattering, with $f$ being any SM fermion
except $e^-$, can be used to look for signatures of presence of such a
$Z'$, both in the neighbourhood of and farther away from the $Z$-pole.
Moreover, we show that one can construct dimensionless
difference-over-sum (DoS) ratios from both cross-section and $A_{FB}$
as well as the corresponding SM predictions. These DoS ratio are
particularly useful observables resilient against the limited
experimental angular coverage of a real detector, where we have
assumed uniform angular acceptance for the detector. The DoS ratio
corresponding to cross-section is well-bounded between $-1$ and $+1$,
and it is sensitive to the presence of a $Z'$-pole around the mass
$m_{Z'}$. The DoS ratio corresponding to $A_{FB}$ is not bounded, but
it is also sensitive to the existence of a $Z'$-pole.

\section*{Acknowldgements}

The work is supported by Basic Science Research Program through the
National Research Foundation of Korea (NRF) funded by the Ministry of
Education (RS-2022-NR070836, RS-2026-25471898).

\end{document}